\documentclass[prl,aps,amssymb,showpacs,twocolumn,superscriptaddress,times,10pt]{revtex4-1}

\usepackage{amsmath}
\usepackage{amssymb}
\usepackage{amsfonts}
\usepackage{listings}
\usepackage{enumerate}
\usepackage{latexsym}
\usepackage{bm}
\usepackage{graphicx}
\usepackage{subfigure}
\usepackage{braket}
\usepackage[pdftex,colorlinks=false]{hyperref}
\usepackage{xcolor}
\usepackage{verbatim}

\usepackage{times}
\usepackage{siunitx}

\newcounter{defcounter}
 \newcommand{\bs}[1]{\ensuremath{\boldsymbol{#1}}}

\begin{document}
\title{Custodial glide symmetry of quantum spin Hall edge modes in WTe$_2$ monolayer}
\author{
Seulgi~Ok}
\address{
 Department of Physics, University of Zurich, Winterthurerstrasse 190, 8057 Zurich, Switzerland
}
\author{
Lukas~Muechler}
\address{
Department of Chemistry, Princeton University, Princeton, New Jersey 08544, USA}

\author{
Domenico Di Sante}
\address{
 Institute for Theoretical Physics and Astrophysics, University of W\"urzburg, Am Hubland, D-97074 W\"urzburg, Germany
}
\author{
Giorgio~Sangiovanni}
\address{
 Institute for Theoretical Physics and Astrophysics, University of W\"urzburg, Am Hubland, D-97074 W\"urzburg, Germany
}
\author{
Ronny~Thomale}
\email{rthomale@physik.uni-wuerzburg.de}
\address{
 Institute for Theoretical Physics and Astrophysics, University of W\"urzburg, Am Hubland, D-97074 W\"urzburg, Germany
}

\author{
Titus~Neupert}
\email{titus.neupert@uzh.ch}
\address{
 Department of Physics, University of Zurich, Winterthurerstrasse 190, 8057 Zurich, Switzerland
}

\begin{abstract}
A monolayer of WTe$_2$ has been shown to display quantum spin Hall (QSH)
edge modes persisting up to 100~K in transport experiments. Based on
density-functional theory calculations and symmetry-based model
building including the role of correlations and substrate support, we develop an effective electronic model for WTe$_2$ which
fundamentally differs from other prototypical QSH
settings: we find that the extraordinary robustness of quantum spin
Hall edge modes in WTe$_2$ roots in a glide symmetry due to which the
topological gap opens away from high-symmetry points in momentum
space. While the indirect bulk gap is much smaller, the glide symmetry implies a large direct gap of up to 1~eV in the Brillouin
zone region of the dispersing edge modes, and hence enables sharply
boundary-localized QSH edge states depending on the specific
boundary orientation.
% ,  We provide a minimal tight-binding model that accurately describes the low-energy physics. From first principles, we study the stability of monolayers of WTe$_2$ on bilayer graphene substrate.
\end{abstract}

\maketitle

\textit{Introduction} --- The quantum spin Hall effect (QSHE) has largely
initiated the era of topological insulators~~\cite{kane2005quantum,bernevig2006quantum, hasan2010colloquium,qi2011topological} and semimetals in
contemporary condensed matter research. Moving beyond its fundamental
relevance as a new quantum state of matter, however, technological applications
can only be brought within reach if QSHE is realized at high operating 
temperatures. After low temperature realizations in HgTe/CdTe quantum
wells~\cite{konig2007quantum,nowack2013imaging,deacon2017josephson} and InAs/GaSb heterostructures~\cite{liu2008quantum,knez2011evidence,du2015robust}, bismuthene has set a new
paradigm for high-temperature QSHE, where the bulk gap is proportional
to twice the atomic spin-orbit coupling of Bi and thus reaches up to
0.8 eV~\cite{reis2017bismuthene,PhysRevB.98.165146}. In many respects, it came as a surprise when a
monolayer of the dichalcogenide WTe$_2$ as yet another material class was subsequently
reported to display QSHE up to 100 K~~\cite{wu2018observation}. There is no evidence
for a particularly large bulk gap in WTe$_2$, and in view of how
significantly the bulk gap in the clean limit is usually reduced to
arrive at the actual gap appearing in transport data, the enormous
robustness of QSHE in WTe$_2$ poses a fundamental quest to
be resolved.

WTe$_2$ is a material whose manifold intricacies arise from its
spin-orbit coupled band structure combined with the W~$d$-orbitals
that hint at electronic correlations~~\cite{qian2014quantum,ali2014large,
soluyanov2015type,keum2015bandgap,jiang2015signature,choe2016understanding,
zheng2016quantum,di2017three,fei2017edge,di2017three,wu2018observation,fatemi2018electrically,fei2018ferroelectric}. 
As a three-dimensional bulk material~\cite{das2016layer,di2017three,wu2017three}, it shows a record-high
magnetoresistance of about a million percent~\cite{ali2014large}. It
has further been predicted to be a Weyl semimetal with strongly
Lorentz symmetry breaking type-II Weyl
cones~\cite{soluyanov2015type}. In the form of monolayers, WTe$_2$ was
already predicted to be a two-dimensional topological
insulator~\cite{qian2014quantum,jiang2015signature,choe2016understanding,zheng2016quantum} 
before it was subsequently confirmed experimentally~\cite{fei2017edge,
  wu2018observation}. Most recently, superconductivity has been observed
in a slightly doped  WTe$_2$ monolayer \cite{fatemi2018electrically}, which
further stresses the potential role of electronic correlations in the compound.
% include a recent study demonstrating the robustness of its
% topological edge modes up to temperatures of about 100~K. This
% should be contrasted with HgTe-CdTe quantum wells, where the quantum
% spin Hall effect is observed up to temperatures of  10~K only. In
% addition, 

In this Letter, we develop an effective low-energy electronic model
for WTe$_2$ monolayers. This is accomplished by different successive
steps. First, we analyze the density functional theory
description and symmetries of WTe$_2$ in light of the existing experimental
evidence, and address how the measured gapped
electronic structure could be rationalized through correlations or
substrate effects. Second, we distill an 8-band model for WTe$_2$
where we identify the dominant atomic and Rashba spin-orbit
terms. As such, our effective model can be conveniently adjusted to fit monolayer WTe$_2$,
and possibly other related QSH materials, for different experimental
setups. Third, we investigate the QSH edge modes in WTe$_2$ for armchair
and zigzag terminations. Due to the glide symmetry of the WTe$_2$
monolayer, the Dirac cones shift away from high-symmetry points and,
in particular for a zigzag termination, allow for the formation of a
big direct gap $\sim 1$ eV protecting the QSH edge modes. Our analysis points towards a
significant termination sensitivity of the edge modes, suggesting
various experimental investigations motivated by our findings.

%Specifically, we show that it is not necessary to resort to stronger correlations to obtain a gapped band structure, but that the inclusion of a substrate can have the same effect. We further provide a minimal effective tight-binding model for the low-energy electronic structure.

\begin{figure}[h]
\centering
\includegraphics[width=0.46\textwidth]{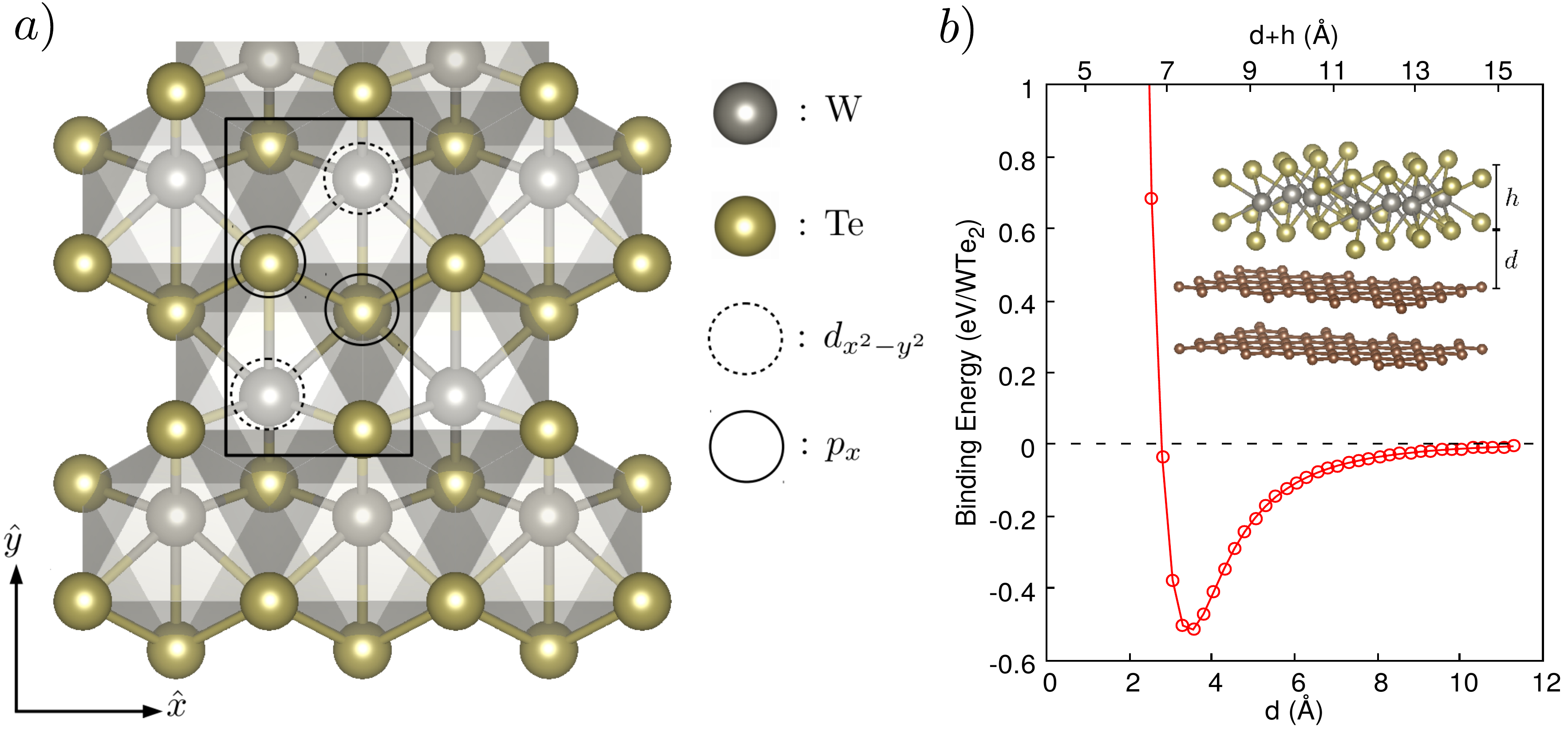}
\caption{a) Lattice structure and unit cell of monolayer
  WTe$_2$. Dashed (solid) circles indicate Wannier functions with
  $d_{x^2-y^2}$ ($p_x$) symmetry contributing to the low-energy
  physics. The other two Te atoms in the unit cell do not play a role
  in the effective low-energy electronic structure. A termination
  parallel to the $\hat{x}$ ($\hat{y}$) direction corresponds to the
  zig-zag (armchair) edge. b) Binding energy of monolayer WTe$_2$
  (height $h$) deposited on a $d$-distant  bilayer graphene (BLG) computed via a van der Waals corrected
  functional. }
%The inset shows a side view of the $2\times
%  2$WTe$_2$/$3\times 6$BLG unit cell we use in our calculations. 
%  }
\label{fig:1qwe}
\end{figure}

{\it Density functional theory analysis} --- The lattice structure of
monolayer WTe$_2$, shown in Fig.~\ref{fig:1qwe}a, is composed of
zig-zag chains of W atoms running along the $\hat{x}$ direction. The
zig-zag nature of the chains endows the lattice with a glide-mirror
symmetry that sends $y\to -y$ combined with half a lattice translation
along the chain. As a common substrate support, WTe$_2$ can be
deposited on top of bilayer graphene (Fig.~\ref{fig:1qwe}b). From a deconstructionist perspective, 
monolayer WTe$_2$ in the absence of spin-orbit coupling would be a
two-dimensional topological semimetal with two gapless Dirac cones that are protected by the non-symmorphic glide-mirror symmetry. 
A band gap arises when spin-orbit coupling is included and gaps out
the Dirac cones. As long as the non-symmorphic symmetry and
time-reversal are preserved, the resulting insulator has to be
topological~\cite{muechler2016topological}.
 In contrast to many other topological insulators, 
%such as graphene (to which the Kane-Mele model
%pertains~\cite{kane2005quantum}),
%bismuthene~\cite{reis2017bismuthene}, and HgTe-CdTe quantum
%wells~\cite{konig2007quantum}, 
the Dirac cones of monolayer WTe$_2$ are not located at  high-symmetry
points in momentum space, but along the glide-mirror line. 
%Here we demonstrate that the extraordinary robustness of the edge modes is directly related to this fact: The localization length of the edge modes is determined by the direct band gap at $k=0$ momentum along the edge, instead of the (indirect) band gap induced by spin-orbit coupling. In WTe$_2$ this direct band gap is of the order of 1~eV, much larger than the spin-orbit coupling strength. As a result, the topological edge modes do not penetrate into the two-dimensional bulk beyond one unit cell, and are therefore less prone to localization.

\begin{figure*}[t]
\centering
\includegraphics[width=0.99\textwidth]{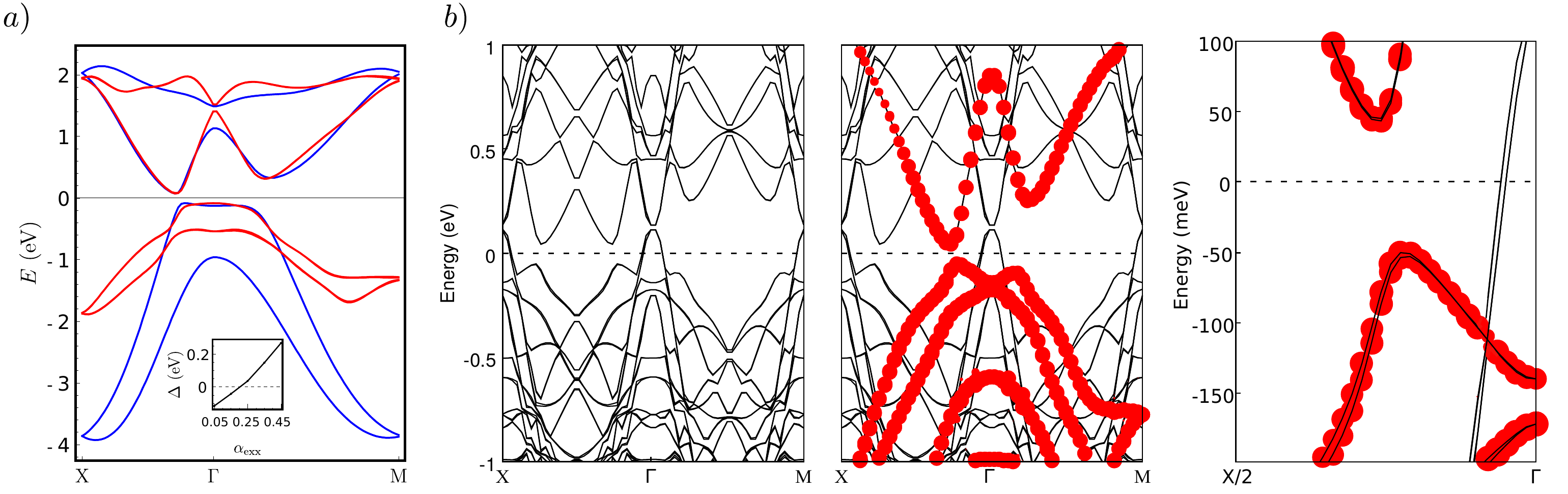}
\caption{a) Comparison of DFT HSE06 band structure (red) and tight-binding band structure (blue) including SOC. The tight-binding model~\eqref{eq:1original} with SOC term of strength $V=0.115$~eV was used, while $\mathcal{H}_{\textrm{R}}^{\textrm{SOC}}$ was neglected.
The inset shows the evolution of the indirect band gap induced by SOC
as a function of the Hartree-Fock exchange. The red curve corresponds
to the HSE06 value $\alpha_{\mathrm{exx}}=0.25$.
b) \textit{Left}: Band structure of monolayer WTe$_2$ (2$\times$2 unit cells) on
BLG (3$\times$6 unit cells) along the high-symmetry lines of the unfolded primitive
Brillouin zone. \textit{Middle}: Same as \textit{left} where the red
circles highlight the unfolding coefficients originating from the
unfolding procedure. These weights are defined from the scalar product
$\langle KN| kn \rangle$, where upper (lower) case symbols $K$ and $N$
($k$ and $n$) identify supercell (primitive cell) momenta and band
indices, and are interpreted as proper spectral weights that allow to
efficiently map the band structure of the primitive cell out of the
band structure of the supercell. A derivation of the unfolding
strategy can be found in
Refs.~\cite{popescu2010effective,ku2010unfolding,tomic2014unfolding}. \textit{Right}:
Zoom around the Dirac cone dispersion. The small splitting of the
bands is due to Rashba SOC.}
\label{fig:2qwe}
\end{figure*}
%\begin{figure*}[t]
%\centering
%\includegraphics[width=0.99\textwidth]{Figures/Figure2.pdf}
%\caption{
%a) Binding energy of monolayer WTe$_2$ on bilayer graphene (BLG) computed by using a standard
%PBE functional (blue points) and a van der Waals corrected functional (red points). The inset shows
%a side view of the $2\times 2$WTe$_2$/$3\times 6$BLG unit cell we used to keep the later strain within
%the values accessible in actual experimental setting. 
%b)--d) Band structure along half of the $\Gamma-X$ high-symmetry line.
%The red circles highlight the unfolding weights.
%}
%\label{fig:2}
%\end{figure*}
The location of the Dirac nodes off high-symmetry points is a challenge for accurate 
first-principle calculations of the low-energy band structure, since slight relative shifts between the energy of the valence band maximum at the $\Gamma$ point and the energy of the Dirac points severely affect the fermiology. 
There is ambiguous experimental evidence with regard to the nature of
the electronic state. Recent angle-resolved photoemission spectroscopy (ARPES) experiments suggest the existence of a fully gapped band structure \cite{tang2017quantum}, while scanning tunneling microscopy (STM) experiments suggest a metallic state \cite{song2018observation}. 
Density functional theory (DFT) calculations based on the generalized
gradient approximation (GGA) predict a type-II Dirac semimetal without
spin-orbit coupling (SOC).
The Dirac cones gap into overlapping
electron and hole pockets once SOC is included, thus still rendering
the resulting state metallic. It is, however, not
clear whether GGA describes the electronic structure
correctly. Rather, the ARPES band structures are well reproduced by
HSE06 calculations of free standing monolayers \cite{tang2017quantum}. This is
in agreement with quantum oscillation experiments on bulk WTe$_2$,
hinting that GGA calculations fail to reproduce the electronic structure close to the Fermi level \cite{rhodes2015role}. 
As an added complication, experiments on the monolayers likely cannot
be directly compared with DFT calculations, since monolayers of
WTe$_2$ are usually realized on substrates. Even though van-der-Waals (vdw)
coupled substrates might only have a moderate effect in terms of
direct hybridization, the substrate could still strain the monolayer, which can induce a transition from a metallic to an
insulating state within GGA calculations. Furthermore, the substrate
could break the glide symmetry of the monolayer,  leading to
other gap opening mechanisms besides SOC. 

\textit{Gap from Fock exchange} --- 
We start out by showing how the inclusion of a certain amount of exact exchange within
DFT leads to a semiconducting ground state and to the opening of a
positive indirect gap in monolayer  WTe$_2$, which would be one way to
address the discrepancy between ARPES experiments and GGA calculations.
%GGA is known to overestimate charge de-localization due to self-interaction errors. 
In WTe$_2$, the tilting of the Dirac cones is controlled by the hoppings along the chain directions and therefore a possible overestimate of these hoppings could account for the metallic band structure predicted by GGA. The inclusion of exact exchange in hybrid-functionals such as HSE06 has been shown to mitigate the effects of self-interaction errors, and therefore is expected to reverse the overlap of the electron and hole pockets along $\Gamma X$. The HSE06 band structure plotted in Fig.~\ref{fig:2qwe}a indeed presents a direct gap~\cite{supp}.
%Here, we systematically study the influence of substrates and post-GGA correlation effects on the direct and indirect band gaps of monolayer WTe$_2$.
The inset of Fig.~\ref{fig:2qwe}a shows the influence of exact exchange on the  indirect band gap, modeled by varying the fraction of exact exchange $\alpha$~\cite{mardirossian2017thirty}. The GGA functional corresponds to $\alpha = 0$, while the HSE06 functional corresponds to $\alpha = 0.25$.
%The direct band gap $\Delta_{d} = E_{con}(\Gamma) - E_{val}(\Gamma)$ is measured from the energy difference of conduction and valence bands at the $\Gamma$ point, while 
The indirect band gap $\Delta = E_{\mathrm{con}}(\bs{k}_c) - E_{\mathrm{val}}(\Gamma)$ is measured from the energy minimum of the conduction band at $\bs{k}_c$ relative to the valence band maximum at $\Gamma$.
%Both direct and indirect band gaps show a linear dependence on the amount of exact exchange. 
The indirect band gap shows a linear dependence on the amount of exact exchange and switches sign from negative to positive at $\alpha \simeq 0.2$, indicating the importance of post-GGA correlation effects in monolayers of WTe$_2$.

\textit{Gap from the substrate} --- 
Strain induced by a substrate can also reduce the hopping strength along the chains in WTe$_2$ and may therefore be an alternative origin for the formation of an indirect gap. 
%In the following, we discuss how the gap formation is strongly favored, and hence does not necessitate the sophisticated HSE06 functional, by explicitly modeling the presence of a substrate.
In actual experimental settings, a  free-standing monolayer WTe$_2$ is grown
on a supporting template. Even though we expect the bonding
to be weak in absolute terms and of vdw nature, the lattice
commensuration plays a crucial role as it induces strain. Bilayer
graphene (BLG) is a typical substrate used to grow monolayers of
transition metal dichalcogenides. In the specific case of WTe$_2$, we
find that a $2\times 2$ reconstruction on BLG $3\times 6$ induces a
lateral tensile strain of $\sim 5.5\%$ along the W zigzag chains. 
In Fig.~\ref{fig:1qwe}b we show the binding energy of WTe$_2$ on BLG
computed by explicitly including the vdw long range interactions. At the
equilibrium distance $d_{\mathrm{eq}}$, the vertical separation between BLG and
the topmost Te layers of WTe$_2$ is $\sim 7.5$~$\textrm{\AA}$, a value that fits well
with the STM measurements reported in~\cite{tang2017quantum}. 

The
band structure, when unfolded in the primitive Brillouin zone
\cite{ku2010unfolding,tomic2014unfolding}, does not show the semimetallic character
typical of free-standing WTe$_2$, but instead is characterized by a
positive indirect gap (Fig.~\ref{fig:2qwe}b). This result originates from
the reduction of the hopping parameters along the W chains, an explicit
consequence of the tensile strain \cite{muechler2016topological}.
Moreover, from the spin splitting of the electronic states around the gapped
Dirac cone, we estimate that the strength of the inversion symmetry
breaking is  $\lesssim 10$ meV. The small magnitude can be attributed to the weak vdw coupling to the substrate.  
%\begin{figure*}[t]
%\centering
%\includegraphics[width=0.99\textwidth]{Figures/Figure3.pdf}
%\caption{
%a)
%Comparison of DFT HSE06 band structure (blue) and tight-binding band structure (red) including SOC.
%The tight-binding model~\eqref{eq:1original} with SOC term of strength $V=0.115$~eV was used.
%b)
%Spectral function defined in~Eq.\eqref{eq:6} showing the topological edge states of tight-binding model~\eqref{eq:1original} on ribbon of 100 sites width with open boundary conditions in $y$ direction. 
%The lower panel shows the localization length $\lambda$ in unit of the lattice constant for the two middle states of the spectrum. Around $k_x=0$, the edge states are localized within a fraction of the unit cell due to the large direct bulk gap. Blue and red curves indicate the upper and the lower middle band, respectively.
%c) 
%Same as b), but for open boundary conditions in the $x$ direction. (Since there is only one edge state at a given momentum, only one curve appears.)
%d)
%Same as b), but including the effect of a Zeeman field. 
%Upper panel: With a Zeeman field parallel to the spin quantization axis defined by the intrinsic SOC, a gap opens in the edge states.
%Lower panel: With a Zeeman field perpendicular to the intrinsic SOC, the edge states remain gapless but the degeneracy point shifts off $k_x=0$.
%In both cases, the magnitude of the Zeeman term was set at 0.05~eV.
%}
%\label{fig:3}
%\end{figure*}

\textit{Effective tight-binding model} --- 
Based on the DFT band structure, we provide the minimal eight-band spin-orbit coupled tight-binding model that has the same spatial symmetries as monolayer WTe$_2$ and quantitatively reproduces its low energy band structure in a window of about 1~eV around the Fermi energy. 
The symmetries of free-standing monolayer WTe$_2$ are time-reversal $T$, a glide mirror $\bar{M}_x$ that sends $x \mapsto -x$ combined with a half lattice translation in $x$ direction, and a two-fold screw symmetry around the $x$ axis $\bar{C}_{2x}$ with the same translation as $\bar{M}_x$. The product of $\bar{M}_x$ and $\bar{C}_{2x}$ is the three-dimensional inversion $\mathcal{I}$. The latter implies a two-fold spin-degeneracy of all bands, and $\bar{M}_{x}$ implies that pairs of these two-fold degenerate bands join into a four-fold degeneracy at $k_x=\pi$. Thus, the minimal insulating band structure with these symmetries has eight bands. 
Building up on the results of~\cite{muechler2016topological}, we choose the corresponding degrees of freedom as spin $s=\uparrow,\downarrow$, sublattice $\kappa=A,B$, and Wannier orbitals $\ell=p,d$. 
We let the Pauli matrices $\sigma_\nu$, $\rho_\nu$, and $\tau_\nu$, for $\nu=0,1,2,3$, act on the $s$, $\kappa$, and $\ell$ degree of freedom, respectively. (Here, $\nu=0$ labels the identity matrices.)
The symmetries are then represented by $T=K \sigma_2\rho_0\tau_0$, mapping $\bs{k}\mapsto -\bs{k}$,
$\bar{M}_x=\sigma_1[\rho_0(1+e^{\mathrm{i}k_x})+\rho_3(1-e^{\mathrm{i}k_x})]\tau_0/2$, mapping $(k_x,k_y)\mapsto (-k_x,k_y)$, and
$\bar{C}_{2x}=\sigma_1[\rho_1(1+e^{\mathrm{i}k_x})+\mathrm{i}\rho_2(1-e^{\mathrm{i}k_x})]\tau_0/2$, mapping $(k_x,k_y)\mapsto (k_x,-k_y)$.
We write the tight-binding Hamiltonian directly in its Bloch
representation in momentum space. Furthermore, we split the model into
a spin-rotation invariant contribution without spin-orbit coupling, an
intrinsic (int) SOC term which still preserves U(1) spin symmetry, and a
Rashba (R) type contribution 
\begin{equation}
\mathcal{H}(\bs{k})=\mathcal{H}_0(\bs{k})+\mathcal{H}^{\textrm{SOC}}_{\textrm{int}}+\mathcal{H}^{\textrm{SOC}}_{\textrm{R}}, 
\label{eq:1original}
\end{equation}
with $\mathcal{H}_0(\bs{k})$, $\mathcal{H}^{\textrm{SOC}}_{\textrm{int}}$, and $\mathcal{H}^{\textrm{SOC}}_{\textrm{R}}$ to be specified below.
%\begin{equation}
%\mathcal{H}_0(\bs{k})=
%\sigma_0
%\otimes
%\left( \begin{array}{cccc}
%\epsilon_d(\bs{k})&0&\tilde{t}_dg_{k_x}e^{\textrm{i} k_y}&\tilde{t}_0f_{k_x}\\
%0&\epsilon_p(\bs{k})&-\tilde{t}_0f_{k_x}&\tilde{t}_pg_{k_x}\\
%\tilde{t}_dg_{k_x}^*e^{-\textrm{i} k_y}&-\tilde{t}_0f_{k_x}^*&\epsilon_d(\bs{k})&0\\
%\tilde{t}_0f_{k_x}^*&\tilde{t}_pg_{k_x}^*&0&\epsilon_p(\bs{k})
%\end{array} \right),
%\label{eq:1}
%\end{equation}
%where $g_{k_x}=1+e^{-\textrm{i} k_x}$, $f_{k_x}=1-e^{-\textrm{i} k_x}$, and 
%$\epsilon_\ell(\bs{k})=\mu_\ell+2 t_\ell \cos k_x+2t_\ell'\cos 2k_x$ for $\ell=p,d$.
%\section{Spinless Hamiltonian}
%We set the tight-binding Hamiltonian of WTe$_2$ in the basis of $\sigma_i \otimes \rho_j \otimes \tau_k$, where $\sigma_i, \rho_j,  \tau_k$ are the $i,j,k$ component of usual Pauli matrix in the spin space, $A$ and $B$ sublattice space, and the sublattice with $d$ and $p$ orbital, respectively.
The explicit form of the Hamiltonian contribution without SOC is 
\begin{equation}
\mathcal{H}_0(\bs{k})=
\sigma_0
\otimes
\left( \begin{array}{cccc}
\epsilon_d(\bs{k})&0&\tilde{t}_dg_{k_x}e^{\textrm{i} k_y}&\tilde{t}_0f_{k_x}\\
0&\epsilon_p(\bs{k})&-\tilde{t}_0f_{k_x}&\tilde{t}_pg_{k_x}\\
\tilde{t}_dg_{k_x}^*e^{-\textrm{i} k_y}&-\tilde{t}_0f_{k_x}^*&\epsilon_d(\bs{k})&0\\
\tilde{t}_0f_{k_x}^*&\tilde{t}_pg_{k_x}^*&0&\epsilon_p(\bs{k})
\end{array} \right),
\label{eq:1}
\end{equation}
where $g_{k_x}=1+e^{-\textrm{i} k_x}$, $f_{k_x}=1-e^{-\textrm{i} k_x}$, and 
$\epsilon_\ell(\bs{k})=\mu_\ell+2 t_\ell \cos k_x+2t_\ell'\cos 2k_x$ for $\ell=p,d$.
By choosing a set of parameters ($\mu_d=0.4935,\mu_p=1.3265,t_d=-0.28,t_d'=0.075,t_p=0.93,t_p'=0.075,t_d^{AB}0.52,t_p^{AB}=0.40,t_0^{AB}=1.02$) in eV units, we get a dispersion which fits well with HSE06 result near the Fermi level.

%\section{Spin-orbit coupled Hamiltonian}
In order to reproduce the gapped dispersion observed in both the case of a substrate or the calculations with HSE06, we include a spin-orbit coupling term that preserves i) TRS, ii) mirror symmetry, and iii) glide symmetry. 
Among a number of possibilities, we particularize to the ones at
lowest order in $\bs{k}=(k_x,k_y)$, in line with our goal to find the
simplest SOC terms.
In the basis we chose in Eq.~\eqref{eq:1}, there are two constant terms that we use in this paper: $\sigma_2 \rho_3 \tau_2$ and $\sigma_3 \rho_3 \tau_2$.
Introducing those, we obtain an intrinsic SOC term of the form
\begin{equation}
\mathcal{H}^{\textrm{SOC}}_{\textrm{int}}=V \sigma_2 \rho_3 \tau_2 + V' \sigma_3 \rho_3 \tau_2,
\label{eq:4}
\end{equation}
with coefficients $V$ and $V'$.
One finds that those two types of terms are related by spin-rotation
by $\pi/4$ around $\sigma_1$, and that the spin-rotation also
satisfies all symmetries, as it is a global unitary transformation of the Hamiltonian.
Therefore, the two terms act equivalently in the Hamiltonian: The
spectrum and the topology of the Hamiltonian with SOC,
Eq.~\eqref{eq:4}, is invariant as long as $V^2 + V'^2$ is kept
unchanged. 
Unless otherwise noted, we assume $V'=0$ throughout the rest of the paper.
% $V,V' \in \mathbb{R}$.
%For this reason we set $V'=0$ without loss of generality so that we have a simpler form
%\begin{equation}
%\mathcal{H}^{\textrm{SOC}}_{\textrm{int}}=V \sigma_2 \rho_3 \tau_2.
%\label{eq:5}
%\end{equation}
We checked that for $V=0.115$~eV, the dispersion in a window of about $\pm0.5\, \mathrm{eV}$ around the Fermi level is very similar to the HSE06 and substrate calculations with SOC (Fig.~\ref{fig:2qwe}a).

\begin{figure*}[t]
\centering
\includegraphics[width=0.99\textwidth]{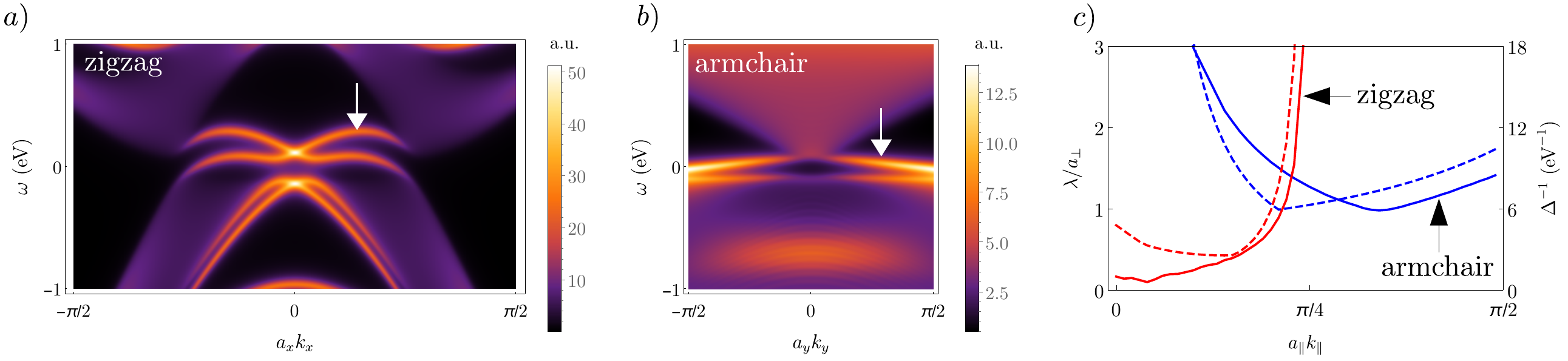}
\caption{a) Spectral function defined in~Eq.\eqref{eq:6} showing the topological edge states of tight-binding model~\eqref{eq:1original} on ribbon of 100 sites width with open boundary conditions in the $y$ direction. b) Same as a), but for open boundary conditions in the $x$ direction.
c) Localization length (solid lines) and inverse gap (dashed lines) as a function of $k_{\parallel}=k_x$ for the case of a), and $k_{\parallel}=k_y$ for the case of b). The relevant bands are highlighted by white arrows in a) and b). 
Around $k_x=0$, the edge states are localized within a fraction of the unit cell due to the large direct bulk gap.}
\label{fig:3qwe}
\end{figure*}

To include the effect of the substrate in the tight-binding model~\eqref{eq:1original}, we add Rashba-type SOC terms $\mathcal{H}^{\textrm{SOC}}_{\textrm{R}}$. 
We choose the ones with the lowest order (constant in $k_x$ and $k_y$) that preserve time-reversal and mirror symmetry, but break the screw symmetry (and thus also break inversion). 
By symmetry considerations similar to those used to derive the intrinsic SOC terms~\eqref{eq:4}, we get $\sigma_2 \rho_0 \tau_2$ and $\sigma_3 \rho_0 \tau_2$ as candidate matrices for the perturbations to the tight-binding Hamiltonian.
One notable feature is that the two candidates are also related by spin-rotation by $\pi/4$ around $\sigma_x$, identically to the intrinsic SOC term.
This fact leads us to a classification depending on whether the intrinsic SOC and the Rashba term are parallel or perpendicular in spin space:
The Rashba term $\sigma_2 \rho_0 \tau_2$ is parallel in spin to (and thus commuting with) the first term in the intrinsic SOC term~\eqref{eq:4}, while it is orthogonal (anticommuting) with the second term.
For the other Rashba term $\sigma_3 \rho_0 \tau_2$ the situation is reversed: It anticommutes with $\sigma_2 \rho_3 \tau_2$ and commutes with $\sigma_3 \rho_3 \tau_2$.
 Furthermore, a Zeeman field $\boldsymbol{B}$ can be added to the tight-binding
 model~\eqref{eq:1original} via the term $\boldsymbol{B} \cdot \boldsymbol{\sigma} \rho_0 \tau_0$.

{\it QSH edge modes ---} Using the value $V=0.115$~eV for the SOC amplitude in the following, we calculate the spectral functions in two different slab geometries
\begin{equation}
A^{\lambda}(\omega,k_{\bar{\lambda}})=\sum_{i} \textrm{Im} \left[ \frac{\langle \psi^{i}(k_{\bar{\lambda}})| P_{\lambda} | \psi^{i}(k_{\bar{\alpha}}) \rangle}{\omega-E^{i}(k_{\bar{\lambda}})+\textrm{i} \delta} \right] ,
\label{eq:6}
\end{equation}
with $\lambda=x,y$ ($\bar{\lambda}=y,x$ for respective case) being the open (periodic) direction of the slab, $P_{\lambda}$ the one-edge projector for $\lambda$-slab, and we used $\delta=0.04$ for the Lorentzian broadening.
The results are plotted in Fig.~\ref{fig:3qwe}a and~\ref{fig:3qwe}b for the zig-zag and armchair edge, respectively. A stark difference between the two edges states is that the former has a Kramers degenerate point in the bulk gap, while the latter does not.
Figure~\ref{fig:3qwe}c shows the localization length of the edge modes
as a function of the momentum parallel to the edge. We observe a
particularly sharp localization of the order of one unit cell for the
zigzag edge near the $k_x=0$ Kramers point. As expected, the
localization length diverges at the momenta where the edge states
connect to the bulk bands. In contrast, the localization length of the
armchair edge states is much larger. For both edge orientations, the
localization lengths correlate well with the inverse difference in
energy between the edge state and the lowest bulk state (see dashed lines in Fig.~\ref{fig:3qwe}c). This identifies the large direct bulk gap around the $\Gamma$ point as the origin of the extreme localization of the zigzag edge modes. 

Furthermore, we investigate the response of the QSH edge modes to an
external Zeeman field. As seen in~\cite{dominguez2018testing}, this magnetic response is
highly sensitive to the type of termination and relative orientation
of the Zeeman field. To leading order, the Zeeman effect opens a gap in the edge states if and only if the field points perpendicular to the spin-quantization axis singled out by the bulk spin-orbit coupling term [$\sigma_2$ for the $V$ term and $\sigma_3$ for the $V'$ term in Eq.~\eqref{eq:4}]~\cite{supp}.

%To include the effect of the substrate in the tight-binding model~\eqref{eq:1original}, we may add Rashba-type SOC terms $\mathcal{H}^{\textrm{SOC}}_{\textrm{R}}$. 
%We choose the ones with the lowest order (constant in $k_x$ and $k_y$) that preserve time-reversal and mirror symmetry, but break the screw symmetry (and thus also break inversion). 
%By symmetry considerations similar to those used to derive the intrinsic SOC terms~\eqref{eq:4}, we get $\sigma_2 \rho_0 \tau_2$ and $\sigma_3 \rho_0 \tau_2$ as candidate matrices for the perturbations to the tight-binding Hamiltonian.
%One interesting feature is that the two candidates are also related by spin-rotation by $\pi/4$ around $\sigma_x$, identically to the intrinsic SOC term.
%This fact leads us to a classification depending on whether the intrinsic SOC and the Rashba term are parallel or perpendicular in spin space:
%The Rashba term $\sigma_2 \rho_0 \tau_2$ is parallel in spin to (and thus commuting with) the first term in the intrinsic SOC term~\eqref{eq:4}, while it is orthogonal (anticommuting) with the second term.
%For the other Rashba term $\sigma_3 \rho_0 \tau_2$ the situation is reversed: It anticommutes with $\sigma_2 \rho_3 \tau_2$ and commutes with $\sigma_3 \rho_3 \tau_2$.

% Furthermore, Zeeman terms can be added to the tight-binding model~\eqref{eq:1original} via the matrices $\sigma_2 \rho_0 \tau_0$ and $\sigma_3 \rho_0 \tau_0$. (See supplemental information for a detailed discussion of the Zeeman effect on the topological boundary states.)

\textit{Discussion} --- Through developing an effective low-energy
electronic model for monolayer WTe$_2$, several directions of
potential experimental and theoretical investigation offer themselves
for further consideration. The motif to employ a glide symmetry to
allow for Dirac cone gap opening away from high-symmetry points is
likely to be applicable to a broad range of materials which so
far have not yet been in the center of attention as
candidates for quantum spin Hall effect. In the specific case of
WTe$_2$, substrate engineering might be intensified to optimize the
electronic setting for a robust QSH phase. Furthermore, even though we
in principle also find a consistent description for WTe$_2$ without
invoking strong electronic correlations, more sophisticated
theoretical approaches may be used to analyze the role of
electron-electron interactions in WTe$_2$. Finally, the high
sensitivity of the QSH edge mode localization length towards the
specific termination can be probed in experiment. In this context, a
rotation of the gate orientation might the most convenient way to
pursue such an investigation.

R.T. thanks R. Cava and L. W. Molenkamp for discussions.
The work was supported by ERC-StG-Thomale-TOPOLECTRICS-336012, ERC-StG-Neupert-757867-PARATOP, DFG-SPP 1666, DFG-SFB 1170, ``ToCoTronics'', and the Swiss National Science Foundation (grant number: 200021\_169061).

%\textit{Note added.} Upon completion of this manuscript, we became
%aware of a contemporaneous effort to perform transport simulations
%based on a suitable effective model for monolayer WeT$_2$...

\bibliography{WTe2}

\begin{thebibliography}{36}%
\makeatletter
\providecommand \@ifxundefined [1]{%
 \@ifx{#1\undefined}
}%
\providecommand \@ifnum [1]{%
 \ifnum #1\expandafter \@firstoftwo
 \else \expandafter \@secondoftwo
 \fi
}%
\providecommand \@ifx [1]{%
 \ifx #1\expandafter \@firstoftwo
 \else \expandafter \@secondoftwo
 \fi
}%
\providecommand \natexlab [1]{#1}%
\providecommand \enquote  [1]{``#1''}%
\providecommand \bibnamefont  [1]{#1}%
\providecommand \bibfnamefont [1]{#1}%
\providecommand \citenamefont [1]{#1}%
\providecommand \href@noop [0]{\@secondoftwo}%
\providecommand \href [0]{\begingroup \@sanitize@url \@href}%
\providecommand \@href[1]{\@@startlink{#1}\@@href}%
\providecommand \@@href[1]{\endgroup#1\@@endlink}%
\providecommand \@sanitize@url [0]{\catcode `\\12\catcode `\$12\catcode
  `\&12\catcode `\#12\catcode `\^12\catcode `\_12\catcode `\%12\relax}%
\providecommand \@@startlink[1]{}%
\providecommand \@@endlink[0]{}%
\providecommand \url  [0]{\begingroup\@sanitize@url \@url }%
\providecommand \@url [1]{\endgroup\@href {#1}{\urlprefix }}%
\providecommand \urlprefix  [0]{URL }%
\providecommand \Eprint [0]{\href }%
\providecommand \doibase [0]{http://dx.doi.org/}%
\providecommand \selectlanguage [0]{\@gobble}%
\providecommand \bibinfo  [0]{\@secondoftwo}%
\providecommand \bibfield  [0]{\@secondoftwo}%
\providecommand \translation [1]{[#1]}%
\providecommand \BibitemOpen [0]{}%
\providecommand \bibitemStop [0]{}%
\providecommand \bibitemNoStop [0]{.\EOS\space}%
\providecommand \EOS [0]{\spacefactor3000\relax}%
\providecommand \BibitemShut  [1]{\csname bibitem#1\endcsname}%
\let\auto@bib@innerbib\@empty
%</preamble>
\bibitem [{\citenamefont {Kane}\ and\ \citenamefont
  {Mele}(2005)}]{kane2005quantum}%
  \BibitemOpen
  \bibfield  {author} {\bibinfo {author} {\bibfnamefont {C.~L.}\ \bibnamefont
  {Kane}}\ and\ \bibinfo {author} {\bibfnamefont {E.~J.}\ \bibnamefont
  {Mele}},\ }\href@noop {} {\bibfield  {journal} {\bibinfo  {journal} {Physical
  Review Letters}\ }\textbf {\bibinfo {volume} {95}},\ \bibinfo {pages}
  {226801} (\bibinfo {year} {2005})}\BibitemShut {NoStop}%
\bibitem [{\citenamefont {Bernevig}\ \emph {et~al.}(2006)\citenamefont
  {Bernevig}, \citenamefont {Hughes},\ and\ \citenamefont
  {Zhang}}]{bernevig2006quantum}%
  \BibitemOpen
  \bibfield  {author} {\bibinfo {author} {\bibfnamefont {B.~A.}\ \bibnamefont
  {Bernevig}}, \bibinfo {author} {\bibfnamefont {T.~L.}\ \bibnamefont
  {Hughes}}, \ and\ \bibinfo {author} {\bibfnamefont {S.-C.}\ \bibnamefont
  {Zhang}},\ }\href@noop {} {\bibfield  {journal} {\bibinfo  {journal}
  {Science}\ }\textbf {\bibinfo {volume} {314}},\ \bibinfo {pages} {1757}
  (\bibinfo {year} {2006})}\BibitemShut {NoStop}%
\bibitem [{\citenamefont {Hasan}\ and\ \citenamefont
  {Kane}(2010)}]{hasan2010colloquium}%
  \BibitemOpen
  \bibfield  {author} {\bibinfo {author} {\bibfnamefont {M.~Z.}\ \bibnamefont
  {Hasan}}\ and\ \bibinfo {author} {\bibfnamefont {C.~L.}\ \bibnamefont
  {Kane}},\ }\href@noop {} {\bibfield  {journal} {\bibinfo  {journal} {Reviews
  of Modern Physics}\ }\textbf {\bibinfo {volume} {82}},\ \bibinfo {pages}
  {3045} (\bibinfo {year} {2010})}\BibitemShut {NoStop}%
\bibitem [{\citenamefont {Qi}\ and\ \citenamefont
  {Zhang}(2011)}]{qi2011topological}%
  \BibitemOpen
  \bibfield  {author} {\bibinfo {author} {\bibfnamefont {X.-L.}\ \bibnamefont
  {Qi}}\ and\ \bibinfo {author} {\bibfnamefont {S.-C.}\ \bibnamefont {Zhang}},\
  }\href@noop {} {\bibfield  {journal} {\bibinfo  {journal} {Reviews of Modern
  Physics}\ }\textbf {\bibinfo {volume} {83}},\ \bibinfo {pages} {1057}
  (\bibinfo {year} {2011})}\BibitemShut {NoStop}%
\bibitem [{\citenamefont {K{\"o}nig}\ \emph {et~al.}(2007)\citenamefont
  {K{\"o}nig}, \citenamefont {Wiedmann}, \citenamefont {Br{\"u}ne},
  \citenamefont {Roth}, \citenamefont {Buhmann}, \citenamefont {Molenkamp},
  \citenamefont {Qi},\ and\ \citenamefont {Zhang}}]{konig2007quantum}%
  \BibitemOpen
  \bibfield  {author} {\bibinfo {author} {\bibfnamefont {M.}~\bibnamefont
  {K{\"o}nig}}, \bibinfo {author} {\bibfnamefont {S.}~\bibnamefont {Wiedmann}},
  \bibinfo {author} {\bibfnamefont {C.}~\bibnamefont {Br{\"u}ne}}, \bibinfo
  {author} {\bibfnamefont {A.}~\bibnamefont {Roth}}, \bibinfo {author}
  {\bibfnamefont {H.}~\bibnamefont {Buhmann}}, \bibinfo {author} {\bibfnamefont
  {L.~W.}\ \bibnamefont {Molenkamp}}, \bibinfo {author} {\bibfnamefont {X.-L.}\
  \bibnamefont {Qi}}, \ and\ \bibinfo {author} {\bibfnamefont {S.-C.}\
  \bibnamefont {Zhang}},\ }\href@noop {} {\bibfield  {journal} {\bibinfo
  {journal} {Science}\ }\textbf {\bibinfo {volume} {318}},\ \bibinfo {pages}
  {766} (\bibinfo {year} {2007})}\BibitemShut {NoStop}%
\bibitem [{\citenamefont {Nowack}\ \emph {et~al.}(2013)\citenamefont {Nowack},
  \citenamefont {Spanton}, \citenamefont {Baenninger}, \citenamefont
  {K{\"o}nig}, \citenamefont {Kirtley}, \citenamefont {Kalisky}, \citenamefont
  {Ames}, \citenamefont {Leubner}, \citenamefont {Br{\"u}ne}, \citenamefont
  {Buhmann} \emph {et~al.}}]{nowack2013imaging}%
  \BibitemOpen
  \bibfield  {author} {\bibinfo {author} {\bibfnamefont {K.~C.}\ \bibnamefont
  {Nowack}}, \bibinfo {author} {\bibfnamefont {E.~M.}\ \bibnamefont {Spanton}},
  \bibinfo {author} {\bibfnamefont {M.}~\bibnamefont {Baenninger}}, \bibinfo
  {author} {\bibfnamefont {M.}~\bibnamefont {K{\"o}nig}}, \bibinfo {author}
  {\bibfnamefont {J.~R.}\ \bibnamefont {Kirtley}}, \bibinfo {author}
  {\bibfnamefont {B.}~\bibnamefont {Kalisky}}, \bibinfo {author} {\bibfnamefont
  {C.}~\bibnamefont {Ames}}, \bibinfo {author} {\bibfnamefont {P.}~\bibnamefont
  {Leubner}}, \bibinfo {author} {\bibfnamefont {C.}~\bibnamefont {Br{\"u}ne}},
  \bibinfo {author} {\bibfnamefont {H.}~\bibnamefont {Buhmann}},  \emph
  {et~al.},\ }\href@noop {} {\bibfield  {journal} {\bibinfo  {journal} {Nature
  Materials}\ }\textbf {\bibinfo {volume} {12}},\ \bibinfo {pages} {787}
  (\bibinfo {year} {2013})}\BibitemShut {NoStop}%
\bibitem [{\citenamefont {Deacon}\ \emph {et~al.}(2017)\citenamefont {Deacon},
  \citenamefont {Wiedenmann}, \citenamefont {Bocquillon}, \citenamefont
  {Dom{\'\i}nguez}, \citenamefont {Klapwijk}, \citenamefont {Leubner},
  \citenamefont {Br{\"u}ne}, \citenamefont {Hankiewicz}, \citenamefont
  {Tarucha}, \citenamefont {Ishibashi} \emph {et~al.}}]{deacon2017josephson}%
  \BibitemOpen
  \bibfield  {author} {\bibinfo {author} {\bibfnamefont {R.~S.}\ \bibnamefont
  {Deacon}}, \bibinfo {author} {\bibfnamefont {J.}~\bibnamefont {Wiedenmann}},
  \bibinfo {author} {\bibfnamefont {E.}~\bibnamefont {Bocquillon}}, \bibinfo
  {author} {\bibfnamefont {F.}~\bibnamefont {Dom{\'\i}nguez}}, \bibinfo
  {author} {\bibfnamefont {T.~M.}\ \bibnamefont {Klapwijk}}, \bibinfo {author}
  {\bibfnamefont {P.}~\bibnamefont {Leubner}}, \bibinfo {author} {\bibfnamefont
  {C.}~\bibnamefont {Br{\"u}ne}}, \bibinfo {author} {\bibfnamefont {E.~M.}\
  \bibnamefont {Hankiewicz}}, \bibinfo {author} {\bibfnamefont
  {S.}~\bibnamefont {Tarucha}}, \bibinfo {author} {\bibfnamefont
  {K.}~\bibnamefont {Ishibashi}},  \emph {et~al.},\ }\href@noop {} {\bibfield
  {journal} {\bibinfo  {journal} {Physical Review X}\ }\textbf {\bibinfo
  {volume} {7}},\ \bibinfo {pages} {021011} (\bibinfo {year}
  {2017})}\BibitemShut {NoStop}%
\bibitem [{\citenamefont {Liu}\ \emph {et~al.}(2008)\citenamefont {Liu},
  \citenamefont {Hughes}, \citenamefont {Qi}, \citenamefont {Wang},\ and\
  \citenamefont {Zhang}}]{liu2008quantum}%
  \BibitemOpen
  \bibfield  {author} {\bibinfo {author} {\bibfnamefont {C.}~\bibnamefont
  {Liu}}, \bibinfo {author} {\bibfnamefont {T.~L.}\ \bibnamefont {Hughes}},
  \bibinfo {author} {\bibfnamefont {X.-L.}\ \bibnamefont {Qi}}, \bibinfo
  {author} {\bibfnamefont {K.}~\bibnamefont {Wang}}, \ and\ \bibinfo {author}
  {\bibfnamefont {S.-C.}\ \bibnamefont {Zhang}},\ }\href@noop {} {\bibfield
  {journal} {\bibinfo  {journal} {Physical Review Letters}\ }\textbf {\bibinfo
  {volume} {100}},\ \bibinfo {pages} {236601} (\bibinfo {year}
  {2008})}\BibitemShut {NoStop}%
\bibitem [{\citenamefont {Knez}\ \emph {et~al.}(2011)\citenamefont {Knez},
  \citenamefont {Du},\ and\ \citenamefont {Sullivan}}]{knez2011evidence}%
  \BibitemOpen
  \bibfield  {author} {\bibinfo {author} {\bibfnamefont {I.}~\bibnamefont
  {Knez}}, \bibinfo {author} {\bibfnamefont {R.-R.}\ \bibnamefont {Du}}, \ and\
  \bibinfo {author} {\bibfnamefont {G.}~\bibnamefont {Sullivan}},\ }\href@noop
  {} {\bibfield  {journal} {\bibinfo  {journal} {Physical Review Letters}\
  }\textbf {\bibinfo {volume} {107}},\ \bibinfo {pages} {136603} (\bibinfo
  {year} {2011})}\BibitemShut {NoStop}%
\bibitem [{\citenamefont {Du}\ \emph {et~al.}(2015)\citenamefont {Du},
  \citenamefont {Knez}, \citenamefont {Sullivan},\ and\ \citenamefont
  {Du}}]{du2015robust}%
  \BibitemOpen
  \bibfield  {author} {\bibinfo {author} {\bibfnamefont {L.}~\bibnamefont
  {Du}}, \bibinfo {author} {\bibfnamefont {I.}~\bibnamefont {Knez}}, \bibinfo
  {author} {\bibfnamefont {G.}~\bibnamefont {Sullivan}}, \ and\ \bibinfo
  {author} {\bibfnamefont {R.-R.}\ \bibnamefont {Du}},\ }\href@noop {}
  {\bibfield  {journal} {\bibinfo  {journal} {Physical Review Letters}\
  }\textbf {\bibinfo {volume} {114}},\ \bibinfo {pages} {096802} (\bibinfo
  {year} {2015})}\BibitemShut {NoStop}%
\bibitem [{\citenamefont {Reis}\ \emph {et~al.}(2017)\citenamefont {Reis},
  \citenamefont {Li}, \citenamefont {Dudy}, \citenamefont {Bauernfeind},
  \citenamefont {Glass}, \citenamefont {Hanke}, \citenamefont {Thomale},
  \citenamefont {Sch{\"a}fer},\ and\ \citenamefont
  {Claessen}}]{reis2017bismuthene}%
  \BibitemOpen
  \bibfield  {author} {\bibinfo {author} {\bibfnamefont {F.}~\bibnamefont
  {Reis}}, \bibinfo {author} {\bibfnamefont {G.}~\bibnamefont {Li}}, \bibinfo
  {author} {\bibfnamefont {L.}~\bibnamefont {Dudy}}, \bibinfo {author}
  {\bibfnamefont {M.}~\bibnamefont {Bauernfeind}}, \bibinfo {author}
  {\bibfnamefont {S.}~\bibnamefont {Glass}}, \bibinfo {author} {\bibfnamefont
  {W.}~\bibnamefont {Hanke}}, \bibinfo {author} {\bibfnamefont
  {R.}~\bibnamefont {Thomale}}, \bibinfo {author} {\bibfnamefont
  {J.}~\bibnamefont {Sch{\"a}fer}}, \ and\ \bibinfo {author} {\bibfnamefont
  {R.}~\bibnamefont {Claessen}},\ }\href@noop {} {\bibfield  {journal}
  {\bibinfo  {journal} {Science}\ }\textbf {\bibinfo {volume} {357}},\ \bibinfo
  {pages} {287} (\bibinfo {year} {2017})}\BibitemShut {NoStop}%
\bibitem [{\citenamefont {Li}\ \emph {et~al.}(2018)\citenamefont {Li},
  \citenamefont {Hanke}, \citenamefont {Hankiewicz}, \citenamefont {Reis},
  \citenamefont {Sch\"afer}, \citenamefont {Claessen}, \citenamefont {Wu},\
  and\ \citenamefont {Thomale}}]{PhysRevB.98.165146}%
  \BibitemOpen
  \bibfield  {author} {\bibinfo {author} {\bibfnamefont {G.}~\bibnamefont
  {Li}}, \bibinfo {author} {\bibfnamefont {W.}~\bibnamefont {Hanke}}, \bibinfo
  {author} {\bibfnamefont {E.~M.}\ \bibnamefont {Hankiewicz}}, \bibinfo
  {author} {\bibfnamefont {F.}~\bibnamefont {Reis}}, \bibinfo {author}
  {\bibfnamefont {J.}~\bibnamefont {Sch\"afer}}, \bibinfo {author}
  {\bibfnamefont {R.}~\bibnamefont {Claessen}}, \bibinfo {author}
  {\bibfnamefont {C.}~\bibnamefont {Wu}}, \ and\ \bibinfo {author}
  {\bibfnamefont {R.}~\bibnamefont {Thomale}},\ }\href {\doibase
  10.1103/PhysRevB.98.165146} {\bibfield  {journal} {\bibinfo  {journal}
  {Physical Review B}\ }\textbf {\bibinfo {volume} {98}},\ \bibinfo {pages}
  {165146} (\bibinfo {year} {2018})}\BibitemShut {NoStop}%
\bibitem [{\citenamefont {Wu}\ \emph {et~al.}(2018)\citenamefont {Wu},
  \citenamefont {Fatemi}, \citenamefont {Gibson}, \citenamefont {Watanabe},
  \citenamefont {Taniguchi}, \citenamefont {Cava},\ and\ \citenamefont
  {Jarillo-Herrero}}]{wu2018observation}%
  \BibitemOpen
  \bibfield  {author} {\bibinfo {author} {\bibfnamefont {S.}~\bibnamefont
  {Wu}}, \bibinfo {author} {\bibfnamefont {V.}~\bibnamefont {Fatemi}}, \bibinfo
  {author} {\bibfnamefont {Q.~D.}\ \bibnamefont {Gibson}}, \bibinfo {author}
  {\bibfnamefont {K.}~\bibnamefont {Watanabe}}, \bibinfo {author}
  {\bibfnamefont {T.}~\bibnamefont {Taniguchi}}, \bibinfo {author}
  {\bibfnamefont {R.~J.}\ \bibnamefont {Cava}}, \ and\ \bibinfo {author}
  {\bibfnamefont {P.}~\bibnamefont {Jarillo-Herrero}},\ }\href@noop {}
  {\bibfield  {journal} {\bibinfo  {journal} {Science}\ }\textbf {\bibinfo
  {volume} {359}},\ \bibinfo {pages} {76} (\bibinfo {year} {2018})}\BibitemShut
  {NoStop}%
\bibitem [{\citenamefont {Qian}\ \emph {et~al.}(2014)\citenamefont {Qian},
  \citenamefont {Liu}, \citenamefont {Fu},\ and\ \citenamefont
  {Li}}]{qian2014quantum}%
  \BibitemOpen
  \bibfield  {author} {\bibinfo {author} {\bibfnamefont {X.}~\bibnamefont
  {Qian}}, \bibinfo {author} {\bibfnamefont {J.}~\bibnamefont {Liu}}, \bibinfo
  {author} {\bibfnamefont {L.}~\bibnamefont {Fu}}, \ and\ \bibinfo {author}
  {\bibfnamefont {J.}~\bibnamefont {Li}},\ }\href@noop {} {\bibfield  {journal}
  {\bibinfo  {journal} {Science}\ }\textbf {\bibinfo {volume} {346}},\ \bibinfo
  {pages} {1344} (\bibinfo {year} {2014})}\BibitemShut {NoStop}%
\bibitem [{\citenamefont {Ali}\ \emph {et~al.}(2014)\citenamefont {Ali},
  \citenamefont {Xiong}, \citenamefont {Flynn}, \citenamefont {Tao},
  \citenamefont {Gibson}, \citenamefont {Schoop}, \citenamefont {Liang},
  \citenamefont {Haldolaarachchige}, \citenamefont {Hirschberger},
  \citenamefont {Ong} \emph {et~al.}}]{ali2014large}%
  \BibitemOpen
  \bibfield  {author} {\bibinfo {author} {\bibfnamefont {M.~N.}\ \bibnamefont
  {Ali}}, \bibinfo {author} {\bibfnamefont {J.}~\bibnamefont {Xiong}}, \bibinfo
  {author} {\bibfnamefont {S.}~\bibnamefont {Flynn}}, \bibinfo {author}
  {\bibfnamefont {J.}~\bibnamefont {Tao}}, \bibinfo {author} {\bibfnamefont
  {Q.~D.}\ \bibnamefont {Gibson}}, \bibinfo {author} {\bibfnamefont {L.~M.}\
  \bibnamefont {Schoop}}, \bibinfo {author} {\bibfnamefont {T.}~\bibnamefont
  {Liang}}, \bibinfo {author} {\bibfnamefont {N.}~\bibnamefont
  {Haldolaarachchige}}, \bibinfo {author} {\bibfnamefont {M.}~\bibnamefont
  {Hirschberger}}, \bibinfo {author} {\bibfnamefont {N.}~\bibnamefont {Ong}},
  \emph {et~al.},\ }\href@noop {} {\bibfield  {journal} {\bibinfo  {journal}
  {Nature}\ }\textbf {\bibinfo {volume} {514}},\ \bibinfo {pages} {205}
  (\bibinfo {year} {2014})}\BibitemShut {NoStop}%
\bibitem [{\citenamefont {Soluyanov}\ \emph {et~al.}(2015)\citenamefont
  {Soluyanov}, \citenamefont {Gresch}, \citenamefont {Wang}, \citenamefont
  {Wu}, \citenamefont {Troyer}, \citenamefont {Dai},\ and\ \citenamefont
  {Bernevig}}]{soluyanov2015type}%
  \BibitemOpen
  \bibfield  {author} {\bibinfo {author} {\bibfnamefont {A.~A.}\ \bibnamefont
  {Soluyanov}}, \bibinfo {author} {\bibfnamefont {D.}~\bibnamefont {Gresch}},
  \bibinfo {author} {\bibfnamefont {Z.}~\bibnamefont {Wang}}, \bibinfo {author}
  {\bibfnamefont {Q.}~\bibnamefont {Wu}}, \bibinfo {author} {\bibfnamefont
  {M.}~\bibnamefont {Troyer}}, \bibinfo {author} {\bibfnamefont
  {X.}~\bibnamefont {Dai}}, \ and\ \bibinfo {author} {\bibfnamefont {B.~A.}\
  \bibnamefont {Bernevig}},\ }\href@noop {} {\bibfield  {journal} {\bibinfo
  {journal} {Nature}\ }\textbf {\bibinfo {volume} {527}},\ \bibinfo {pages}
  {495} (\bibinfo {year} {2015})}\BibitemShut {NoStop}%
\bibitem [{\citenamefont {Keum}\ \emph {et~al.}(2015)\citenamefont {Keum},
  \citenamefont {Cho}, \citenamefont {Kim}, \citenamefont {Choe}, \citenamefont
  {Sung}, \citenamefont {Kan}, \citenamefont {Kang}, \citenamefont {Hwang},
  \citenamefont {Kim}, \citenamefont {Yang} \emph {et~al.}}]{keum2015bandgap}%
  \BibitemOpen
  \bibfield  {author} {\bibinfo {author} {\bibfnamefont {D.~H.}\ \bibnamefont
  {Keum}}, \bibinfo {author} {\bibfnamefont {S.}~\bibnamefont {Cho}}, \bibinfo
  {author} {\bibfnamefont {J.~H.}\ \bibnamefont {Kim}}, \bibinfo {author}
  {\bibfnamefont {D.-H.}\ \bibnamefont {Choe}}, \bibinfo {author}
  {\bibfnamefont {H.-J.}\ \bibnamefont {Sung}}, \bibinfo {author}
  {\bibfnamefont {M.}~\bibnamefont {Kan}}, \bibinfo {author} {\bibfnamefont
  {H.}~\bibnamefont {Kang}}, \bibinfo {author} {\bibfnamefont {J.-Y.}\
  \bibnamefont {Hwang}}, \bibinfo {author} {\bibfnamefont {S.~W.}\ \bibnamefont
  {Kim}}, \bibinfo {author} {\bibfnamefont {H.}~\bibnamefont {Yang}},  \emph
  {et~al.},\ }\href@noop {} {\bibfield  {journal} {\bibinfo  {journal} {Nature
  Physics}\ }\textbf {\bibinfo {volume} {11}},\ \bibinfo {pages} {482}
  (\bibinfo {year} {2015})}\BibitemShut {NoStop}%
\bibitem [{\citenamefont {Jiang}\ \emph {et~al.}(2015)\citenamefont {Jiang},
  \citenamefont {Tang}, \citenamefont {Pan}, \citenamefont {Liu}, \citenamefont
  {Niu}, \citenamefont {Wang}, \citenamefont {Xu}, \citenamefont {Yang},
  \citenamefont {Xie}, \citenamefont {Song} \emph
  {et~al.}}]{jiang2015signature}%
  \BibitemOpen
  \bibfield  {author} {\bibinfo {author} {\bibfnamefont {J.}~\bibnamefont
  {Jiang}}, \bibinfo {author} {\bibfnamefont {F.}~\bibnamefont {Tang}},
  \bibinfo {author} {\bibfnamefont {X.}~\bibnamefont {Pan}}, \bibinfo {author}
  {\bibfnamefont {H.}~\bibnamefont {Liu}}, \bibinfo {author} {\bibfnamefont
  {X.}~\bibnamefont {Niu}}, \bibinfo {author} {\bibfnamefont {Y.}~\bibnamefont
  {Wang}}, \bibinfo {author} {\bibfnamefont {D.}~\bibnamefont {Xu}}, \bibinfo
  {author} {\bibfnamefont {H.}~\bibnamefont {Yang}}, \bibinfo {author}
  {\bibfnamefont {B.}~\bibnamefont {Xie}}, \bibinfo {author} {\bibfnamefont
  {F.}~\bibnamefont {Song}},  \emph {et~al.},\ }\href@noop {} {\bibfield
  {journal} {\bibinfo  {journal} {Physical Review Letters}\ }\textbf {\bibinfo
  {volume} {115}},\ \bibinfo {pages} {166601} (\bibinfo {year}
  {2015})}\BibitemShut {NoStop}%
\bibitem [{\citenamefont {Choe}\ \emph {et~al.}(2016)\citenamefont {Choe},
  \citenamefont {Sung},\ and\ \citenamefont {Chang}}]{choe2016understanding}%
  \BibitemOpen
  \bibfield  {author} {\bibinfo {author} {\bibfnamefont {D.-H.}\ \bibnamefont
  {Choe}}, \bibinfo {author} {\bibfnamefont {H.-J.}\ \bibnamefont {Sung}}, \
  and\ \bibinfo {author} {\bibfnamefont {K.~J.}\ \bibnamefont {Chang}},\
  }\href@noop {} {\bibfield  {journal} {\bibinfo  {journal} {Physical Review
  B}\ }\textbf {\bibinfo {volume} {93}},\ \bibinfo {pages} {125109} (\bibinfo
  {year} {2016})}\BibitemShut {NoStop}%
\bibitem [{\citenamefont {Zheng}\ \emph {et~al.}(2016)\citenamefont {Zheng},
  \citenamefont {Cai}, \citenamefont {Ge}, \citenamefont {Zhang}, \citenamefont
  {Liu}, \citenamefont {Lu}, \citenamefont {Zhang}, \citenamefont {Qiu},
  \citenamefont {Taniguchi}, \citenamefont {Watanabe} \emph
  {et~al.}}]{zheng2016quantum}%
  \BibitemOpen
  \bibfield  {author} {\bibinfo {author} {\bibfnamefont {F.}~\bibnamefont
  {Zheng}}, \bibinfo {author} {\bibfnamefont {C.}~\bibnamefont {Cai}}, \bibinfo
  {author} {\bibfnamefont {S.}~\bibnamefont {Ge}}, \bibinfo {author}
  {\bibfnamefont {X.}~\bibnamefont {Zhang}}, \bibinfo {author} {\bibfnamefont
  {X.}~\bibnamefont {Liu}}, \bibinfo {author} {\bibfnamefont {H.}~\bibnamefont
  {Lu}}, \bibinfo {author} {\bibfnamefont {Y.}~\bibnamefont {Zhang}}, \bibinfo
  {author} {\bibfnamefont {J.}~\bibnamefont {Qiu}}, \bibinfo {author}
  {\bibfnamefont {T.}~\bibnamefont {Taniguchi}}, \bibinfo {author}
  {\bibfnamefont {K.}~\bibnamefont {Watanabe}},  \emph {et~al.},\ }\href@noop
  {} {\bibfield  {journal} {\bibinfo  {journal} {Advanced Materials}\ }\textbf
  {\bibinfo {volume} {28}},\ \bibinfo {pages} {4845} (\bibinfo {year}
  {2016})}\BibitemShut {NoStop}%
\bibitem [{\citenamefont {Di~Sante}\ \emph {et~al.}(2017)\citenamefont
  {Di~Sante}, \citenamefont {Das}, \citenamefont {Bigi}, \citenamefont
  {Erg{\"o}nenc}, \citenamefont {G{\"u}rtler}, \citenamefont {Krieger},
  \citenamefont {Schmitt}, \citenamefont {Ali}, \citenamefont {Rossi},
  \citenamefont {Thomale} \emph {et~al.}}]{di2017three}%
  \BibitemOpen
  \bibfield  {author} {\bibinfo {author} {\bibfnamefont {D.}~\bibnamefont
  {Di~Sante}}, \bibinfo {author} {\bibfnamefont {P.~K.}\ \bibnamefont {Das}},
  \bibinfo {author} {\bibfnamefont {C.}~\bibnamefont {Bigi}}, \bibinfo {author}
  {\bibfnamefont {Z.}~\bibnamefont {Erg{\"o}nenc}}, \bibinfo {author}
  {\bibfnamefont {N.}~\bibnamefont {G{\"u}rtler}}, \bibinfo {author}
  {\bibfnamefont {J.}~\bibnamefont {Krieger}}, \bibinfo {author} {\bibfnamefont
  {T.}~\bibnamefont {Schmitt}}, \bibinfo {author} {\bibfnamefont
  {M.}~\bibnamefont {Ali}}, \bibinfo {author} {\bibfnamefont {G.}~\bibnamefont
  {Rossi}}, \bibinfo {author} {\bibfnamefont {R.}~\bibnamefont {Thomale}},
  \emph {et~al.},\ }\href@noop {} {\bibfield  {journal} {\bibinfo  {journal}
  {Physical Review Letters}\ }\textbf {\bibinfo {volume} {119}},\ \bibinfo
  {pages} {026403} (\bibinfo {year} {2017})}\BibitemShut {NoStop}%
\bibitem [{\citenamefont {Fei}\ \emph {et~al.}(2017)\citenamefont {Fei},
  \citenamefont {Palomaki}, \citenamefont {Wu}, \citenamefont {Zhao},
  \citenamefont {Cai}, \citenamefont {Sun}, \citenamefont {Nguyen},
  \citenamefont {Finney}, \citenamefont {Xu},\ and\ \citenamefont
  {Cobden}}]{fei2017edge}%
  \BibitemOpen
  \bibfield  {author} {\bibinfo {author} {\bibfnamefont {Z.}~\bibnamefont
  {Fei}}, \bibinfo {author} {\bibfnamefont {T.}~\bibnamefont {Palomaki}},
  \bibinfo {author} {\bibfnamefont {S.}~\bibnamefont {Wu}}, \bibinfo {author}
  {\bibfnamefont {W.}~\bibnamefont {Zhao}}, \bibinfo {author} {\bibfnamefont
  {X.}~\bibnamefont {Cai}}, \bibinfo {author} {\bibfnamefont {B.}~\bibnamefont
  {Sun}}, \bibinfo {author} {\bibfnamefont {P.}~\bibnamefont {Nguyen}},
  \bibinfo {author} {\bibfnamefont {J.}~\bibnamefont {Finney}}, \bibinfo
  {author} {\bibfnamefont {X.}~\bibnamefont {Xu}}, \ and\ \bibinfo {author}
  {\bibfnamefont {D.~H.}\ \bibnamefont {Cobden}},\ }\href@noop {} {\bibfield
  {journal} {\bibinfo  {journal} {Nature Physics}\ }\textbf {\bibinfo {volume}
  {13}},\ \bibinfo {pages} {677} (\bibinfo {year} {2017})}\BibitemShut
  {NoStop}%
\bibitem [{\citenamefont {Fatemi}\ \emph {et~al.}(2018)\citenamefont {Fatemi},
  \citenamefont {Wu}, \citenamefont {Cao}, \citenamefont {Bretheau},
  \citenamefont {Gibson}, \citenamefont {Watanabe}, \citenamefont {Taniguchi},
  \citenamefont {Cava},\ and\ \citenamefont
  {Jarillo-Herrero}}]{fatemi2018electrically}%
  \BibitemOpen
  \bibfield  {author} {\bibinfo {author} {\bibfnamefont {V.}~\bibnamefont
  {Fatemi}}, \bibinfo {author} {\bibfnamefont {S.}~\bibnamefont {Wu}}, \bibinfo
  {author} {\bibfnamefont {Y.}~\bibnamefont {Cao}}, \bibinfo {author}
  {\bibfnamefont {L.}~\bibnamefont {Bretheau}}, \bibinfo {author}
  {\bibfnamefont {Q.~D.}\ \bibnamefont {Gibson}}, \bibinfo {author}
  {\bibfnamefont {K.}~\bibnamefont {Watanabe}}, \bibinfo {author}
  {\bibfnamefont {T.}~\bibnamefont {Taniguchi}}, \bibinfo {author}
  {\bibfnamefont {R.~J.}\ \bibnamefont {Cava}}, \ and\ \bibinfo {author}
  {\bibfnamefont {P.}~\bibnamefont {Jarillo-Herrero}},\ }\href@noop {}
  {\bibfield  {journal} {\bibinfo  {journal} {Science}\ }\textbf {\bibinfo
  {volume} {10}},\ \bibinfo {pages} {1126} (\bibinfo {year}
  {2018})}\BibitemShut {NoStop}%
\bibitem [{\citenamefont {Fei}\ \emph {et~al.}(2018)\citenamefont {Fei},
  \citenamefont {Zhao}, \citenamefont {Palomaki}, \citenamefont {Sun},
  \citenamefont {Miller}, \citenamefont {Zhao}, \citenamefont {Yan},
  \citenamefont {Xu},\ and\ \citenamefont {Cobden}}]{fei2018ferroelectric}%
  \BibitemOpen
  \bibfield  {author} {\bibinfo {author} {\bibfnamefont {Z.}~\bibnamefont
  {Fei}}, \bibinfo {author} {\bibfnamefont {W.}~\bibnamefont {Zhao}}, \bibinfo
  {author} {\bibfnamefont {T.~A.}\ \bibnamefont {Palomaki}}, \bibinfo {author}
  {\bibfnamefont {B.}~\bibnamefont {Sun}}, \bibinfo {author} {\bibfnamefont
  {M.~K.}\ \bibnamefont {Miller}}, \bibinfo {author} {\bibfnamefont
  {Z.}~\bibnamefont {Zhao}}, \bibinfo {author} {\bibfnamefont {J.}~\bibnamefont
  {Yan}}, \bibinfo {author} {\bibfnamefont {X.}~\bibnamefont {Xu}}, \ and\
  \bibinfo {author} {\bibfnamefont {D.~H.}\ \bibnamefont {Cobden}},\
  }\href@noop {} {\bibfield  {journal} {\bibinfo  {journal} {Nature}\ }\textbf
  {\bibinfo {volume} {560}},\ \bibinfo {pages} {336} (\bibinfo {year}
  {2018})}\BibitemShut {NoStop}%
\bibitem [{\citenamefont {Das}\ \emph {et~al.}(2016)\citenamefont {Das},
  \citenamefont {Di~Sante}, \citenamefont {Vobornik}, \citenamefont {Fujii},
  \citenamefont {Okuda}, \citenamefont {Bruyer}, \citenamefont {Gyenis},
  \citenamefont {Feldman}, \citenamefont {Tao}, \citenamefont {Ciancio} \emph
  {et~al.}}]{das2016layer}%
  \BibitemOpen
  \bibfield  {author} {\bibinfo {author} {\bibfnamefont {P.~K.}\ \bibnamefont
  {Das}}, \bibinfo {author} {\bibfnamefont {D.}~\bibnamefont {Di~Sante}},
  \bibinfo {author} {\bibfnamefont {I.}~\bibnamefont {Vobornik}}, \bibinfo
  {author} {\bibfnamefont {J.}~\bibnamefont {Fujii}}, \bibinfo {author}
  {\bibfnamefont {T.}~\bibnamefont {Okuda}}, \bibinfo {author} {\bibfnamefont
  {E.}~\bibnamefont {Bruyer}}, \bibinfo {author} {\bibfnamefont
  {A.}~\bibnamefont {Gyenis}}, \bibinfo {author} {\bibfnamefont
  {B.}~\bibnamefont {Feldman}}, \bibinfo {author} {\bibfnamefont
  {J.}~\bibnamefont {Tao}}, \bibinfo {author} {\bibfnamefont {R.}~\bibnamefont
  {Ciancio}},  \emph {et~al.},\ }\href@noop {} {\bibfield  {journal} {\bibinfo
  {journal} {Nature Communications}\ }\textbf {\bibinfo {volume} {7}},\
  \bibinfo {pages} {10847} (\bibinfo {year} {2016})}\BibitemShut {NoStop}%
\bibitem [{\citenamefont {Wu}\ \emph {et~al.}(2017)\citenamefont {Wu},
  \citenamefont {Jo}, \citenamefont {Mou}, \citenamefont {Huang}, \citenamefont
  {Bud'ko}, \citenamefont {Canfield},\ and\ \citenamefont
  {Kaminski}}]{wu2017three}%
  \BibitemOpen
  \bibfield  {author} {\bibinfo {author} {\bibfnamefont {Y.}~\bibnamefont
  {Wu}}, \bibinfo {author} {\bibfnamefont {N.~H.}\ \bibnamefont {Jo}}, \bibinfo
  {author} {\bibfnamefont {D.}~\bibnamefont {Mou}}, \bibinfo {author}
  {\bibfnamefont {L.}~\bibnamefont {Huang}}, \bibinfo {author} {\bibfnamefont
  {S.~L.}\ \bibnamefont {Bud'ko}}, \bibinfo {author} {\bibfnamefont {P.~C.}\
  \bibnamefont {Canfield}}, \ and\ \bibinfo {author} {\bibfnamefont
  {A.}~\bibnamefont {Kaminski}},\ }\href@noop {} {\bibfield  {journal}
  {\bibinfo  {journal} {Physical Review B}\ }\textbf {\bibinfo {volume} {95}},\
  \bibinfo {pages} {195138} (\bibinfo {year} {2017})}\BibitemShut {NoStop}%
\bibitem [{\citenamefont {Muechler}\ \emph {et~al.}(2016)\citenamefont
  {Muechler}, \citenamefont {Alexandradinata}, \citenamefont {Neupert},\ and\
  \citenamefont {Car}}]{muechler2016topological}%
  \BibitemOpen
  \bibfield  {author} {\bibinfo {author} {\bibfnamefont {L.}~\bibnamefont
  {Muechler}}, \bibinfo {author} {\bibfnamefont {A.}~\bibnamefont
  {Alexandradinata}}, \bibinfo {author} {\bibfnamefont {T.}~\bibnamefont
  {Neupert}}, \ and\ \bibinfo {author} {\bibfnamefont {R.}~\bibnamefont
  {Car}},\ }\href@noop {} {\bibfield  {journal} {\bibinfo  {journal} {Physical
  Review X}\ }\textbf {\bibinfo {volume} {6}},\ \bibinfo {pages} {041069}
  (\bibinfo {year} {2016})}\BibitemShut {NoStop}%
\bibitem [{\citenamefont {Popescu}\ and\ \citenamefont
  {Zunger}(2010)}]{popescu2010effective}%
  \BibitemOpen
  \bibfield  {author} {\bibinfo {author} {\bibfnamefont {V.}~\bibnamefont
  {Popescu}}\ and\ \bibinfo {author} {\bibfnamefont {A.}~\bibnamefont
  {Zunger}},\ }\href@noop {} {\bibfield  {journal} {\bibinfo  {journal}
  {Physical Review Letters}\ }\textbf {\bibinfo {volume} {104}},\ \bibinfo
  {pages} {236403} (\bibinfo {year} {2010})}\BibitemShut {NoStop}%
\bibitem [{\citenamefont {Ku}\ \emph {et~al.}(2010)\citenamefont {Ku},
  \citenamefont {Berlijn}, \citenamefont {Lee} \emph
  {et~al.}}]{ku2010unfolding}%
  \BibitemOpen
  \bibfield  {author} {\bibinfo {author} {\bibfnamefont {W.}~\bibnamefont
  {Ku}}, \bibinfo {author} {\bibfnamefont {T.}~\bibnamefont {Berlijn}},
  \bibinfo {author} {\bibfnamefont {C.-C.}\ \bibnamefont {Lee}},  \emph
  {et~al.},\ }\href@noop {} {\bibfield  {journal} {\bibinfo  {journal}
  {Physical Review Letters}\ }\textbf {\bibinfo {volume} {104}},\ \bibinfo
  {pages} {216401} (\bibinfo {year} {2010})}\BibitemShut {NoStop}%
\bibitem [{\citenamefont {Tomi{\'c}}\ \emph {et~al.}(2014)\citenamefont
  {Tomi{\'c}}, \citenamefont {Jeschke},\ and\ \citenamefont
  {Valent{\'\i}}}]{tomic2014unfolding}%
  \BibitemOpen
  \bibfield  {author} {\bibinfo {author} {\bibfnamefont {M.}~\bibnamefont
  {Tomi{\'c}}}, \bibinfo {author} {\bibfnamefont {H.~O.}\ \bibnamefont
  {Jeschke}}, \ and\ \bibinfo {author} {\bibfnamefont {R.}~\bibnamefont
  {Valent{\'\i}}},\ }\href@noop {} {\bibfield  {journal} {\bibinfo  {journal}
  {Physical Review B}\ }\textbf {\bibinfo {volume} {90}},\ \bibinfo {pages}
  {195121} (\bibinfo {year} {2014})}\BibitemShut {NoStop}%
\bibitem [{\citenamefont {Tang}\ \emph {et~al.}(2017)\citenamefont {Tang},
  \citenamefont {Zhang}, \citenamefont {Wong}, \citenamefont {Pedramrazi},
  \citenamefont {Tsai}, \citenamefont {Jia}, \citenamefont {Moritz},
  \citenamefont {Claassen}, \citenamefont {Ryu}, \citenamefont {Kahn} \emph
  {et~al.}}]{tang2017quantum}%
  \BibitemOpen
  \bibfield  {author} {\bibinfo {author} {\bibfnamefont {S.}~\bibnamefont
  {Tang}}, \bibinfo {author} {\bibfnamefont {C.}~\bibnamefont {Zhang}},
  \bibinfo {author} {\bibfnamefont {D.}~\bibnamefont {Wong}}, \bibinfo {author}
  {\bibfnamefont {Z.}~\bibnamefont {Pedramrazi}}, \bibinfo {author}
  {\bibfnamefont {H.-Z.}\ \bibnamefont {Tsai}}, \bibinfo {author}
  {\bibfnamefont {C.}~\bibnamefont {Jia}}, \bibinfo {author} {\bibfnamefont
  {B.}~\bibnamefont {Moritz}}, \bibinfo {author} {\bibfnamefont
  {M.}~\bibnamefont {Claassen}}, \bibinfo {author} {\bibfnamefont
  {H.}~\bibnamefont {Ryu}}, \bibinfo {author} {\bibfnamefont {S.}~\bibnamefont
  {Kahn}},  \emph {et~al.},\ }\href@noop {} {\bibfield  {journal} {\bibinfo
  {journal} {Nature Physics}\ }\textbf {\bibinfo {volume} {13}},\ \bibinfo
  {pages} {683} (\bibinfo {year} {2017})}\BibitemShut {NoStop}%
\bibitem [{\citenamefont {Song}\ \emph {et~al.}(2018)\citenamefont {Song},
  \citenamefont {Jia}, \citenamefont {Zhang}, \citenamefont {Zhu},
  \citenamefont {Shi}, \citenamefont {Wang}, \citenamefont {Zhu}, \citenamefont
  {Yuan}, \citenamefont {Zhang}, \citenamefont {Xing} \emph
  {et~al.}}]{song2018observation}%
  \BibitemOpen
  \bibfield  {author} {\bibinfo {author} {\bibfnamefont {Y.-H.}\ \bibnamefont
  {Song}}, \bibinfo {author} {\bibfnamefont {Z.-Y.}\ \bibnamefont {Jia}},
  \bibinfo {author} {\bibfnamefont {D.}~\bibnamefont {Zhang}}, \bibinfo
  {author} {\bibfnamefont {X.-Y.}\ \bibnamefont {Zhu}}, \bibinfo {author}
  {\bibfnamefont {Z.-Q.}\ \bibnamefont {Shi}}, \bibinfo {author} {\bibfnamefont
  {H.}~\bibnamefont {Wang}}, \bibinfo {author} {\bibfnamefont {L.}~\bibnamefont
  {Zhu}}, \bibinfo {author} {\bibfnamefont {Q.-Q.}\ \bibnamefont {Yuan}},
  \bibinfo {author} {\bibfnamefont {H.}~\bibnamefont {Zhang}}, \bibinfo
  {author} {\bibfnamefont {D.-Y.}\ \bibnamefont {Xing}},  \emph {et~al.},\
  }\href@noop {} {\bibfield  {journal} {\bibinfo  {journal} {Nature
  Communications}\ }\textbf {\bibinfo {volume} {9}},\ \bibinfo {pages} {4071}
  (\bibinfo {year} {2018})}\BibitemShut {NoStop}%
\bibitem [{\citenamefont {Rhodes}\ \emph {et~al.}(2015)\citenamefont {Rhodes},
  \citenamefont {Das}, \citenamefont {Zhang}, \citenamefont {Zeng},
  \citenamefont {Pradhan}, \citenamefont {Kikugawa}, \citenamefont
  {Manousakis},\ and\ \citenamefont {Balicas}}]{rhodes2015role}%
  \BibitemOpen
  \bibfield  {author} {\bibinfo {author} {\bibfnamefont {D.}~\bibnamefont
  {Rhodes}}, \bibinfo {author} {\bibfnamefont {S.}~\bibnamefont {Das}},
  \bibinfo {author} {\bibfnamefont {Q.~R.}\ \bibnamefont {Zhang}}, \bibinfo
  {author} {\bibfnamefont {B.}~\bibnamefont {Zeng}}, \bibinfo {author}
  {\bibfnamefont {N.}~\bibnamefont {Pradhan}}, \bibinfo {author} {\bibfnamefont
  {N.}~\bibnamefont {Kikugawa}}, \bibinfo {author} {\bibfnamefont
  {E.}~\bibnamefont {Manousakis}}, \ and\ \bibinfo {author} {\bibfnamefont
  {L.}~\bibnamefont {Balicas}},\ }\href@noop {} {\bibfield  {journal} {\bibinfo
   {journal} {Physical Review B}\ }\textbf {\bibinfo {volume} {92}},\ \bibinfo
  {pages} {125152} (\bibinfo {year} {2015})}\BibitemShut {NoStop}%
\bibitem [{sup()}]{supp}%
  \BibitemOpen
  \href@noop {} {}\bibinfo {note} {Refer to the Supplemental
  Material.}\BibitemShut {Stop}%
\bibitem [{\citenamefont {Mardirossian}\ and\ \citenamefont
  {Head-Gordon}(2017)}]{mardirossian2017thirty}%
  \BibitemOpen
  \bibfield  {author} {\bibinfo {author} {\bibfnamefont {N.}~\bibnamefont
  {Mardirossian}}\ and\ \bibinfo {author} {\bibfnamefont {M.}~\bibnamefont
  {Head-Gordon}},\ }\href@noop {} {\bibfield  {journal} {\bibinfo  {journal}
  {Molecular Physics}\ }\textbf {\bibinfo {volume} {115}},\ \bibinfo {pages}
  {2315} (\bibinfo {year} {2017})}\BibitemShut {NoStop}%
\bibitem [{\citenamefont {Dominguez}\ \emph {et~al.}(2018)\citenamefont
  {Dominguez}, \citenamefont {Scharf}, \citenamefont {Li}, \citenamefont
  {Sch\"afer}, \citenamefont {Claessen}, \citenamefont {Hanke}, \citenamefont
  {Thomale},\ and\ \citenamefont {Hankiewicz}}]{dominguez2018testing}%
  \BibitemOpen
  \bibfield  {author} {\bibinfo {author} {\bibfnamefont {F.}~\bibnamefont
  {Dominguez}}, \bibinfo {author} {\bibfnamefont {B.}~\bibnamefont {Scharf}},
  \bibinfo {author} {\bibfnamefont {G.}~\bibnamefont {Li}}, \bibinfo {author}
  {\bibfnamefont {J.}~\bibnamefont {Sch\"afer}}, \bibinfo {author}
  {\bibfnamefont {R.}~\bibnamefont {Claessen}}, \bibinfo {author}
  {\bibfnamefont {W.}~\bibnamefont {Hanke}}, \bibinfo {author} {\bibfnamefont
  {R.}~\bibnamefont {Thomale}}, \ and\ \bibinfo {author} {\bibfnamefont
  {E.~M.}\ \bibnamefont {Hankiewicz}},\ }\href {\doibase
  10.1103/PhysRevB.98.161407} {\bibfield  {journal} {\bibinfo  {journal}
  {Physical Review B}\ }\textbf {\bibinfo {volume} {98}},\ \bibinfo {pages}
  {161407} (\bibinfo {year} {2018})}\BibitemShut {NoStop}%
\end{thebibliography}%
\clearpage
\begin{widetext}
\section*{Supplementary Information}

\subsection{Effect of Zeeman field on the edge states}
\begin{figure}[h]
\centering
\includegraphics[width=0.98\textwidth]{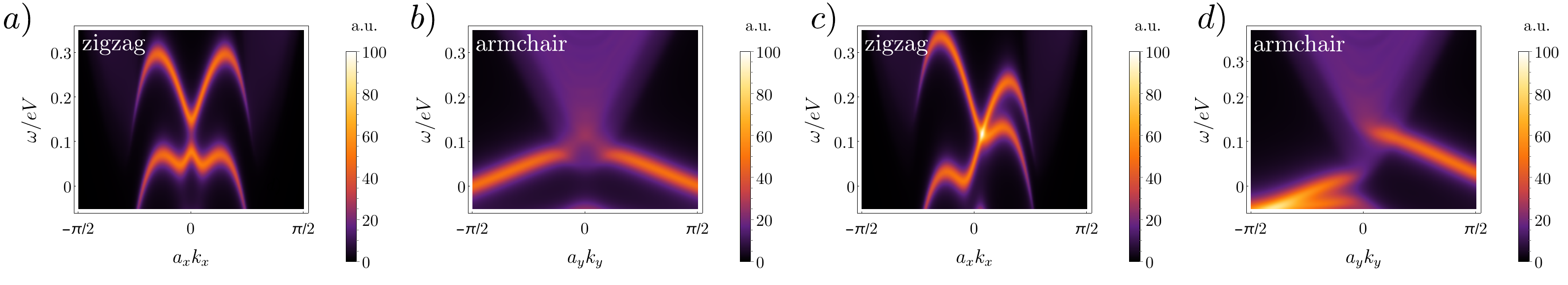}
\caption{Calculation results of spectral function \eqref{eq:6} with the effect of a Zeeman field at the magnitude 0.05~eV. The Zeeman field is perpendicular (parallel) to the spin quantization axis defined by the intrinsic SOC in the left (right) two panels, resulting in the gap-opening (tilting) of the middle states. a) and c) is made in a ribbon geometry open in the $y$ direction, whereas b) and d) is open in the $x$ direction.}
\label{fig:supple2qwe}
\end{figure}

To study how a Zeeman field affects the topological edge states of WTe$_2$, we study the Hamiltonian
\begin{equation}
\widetilde{\mathcal{H}} (\bs{k})=\mathcal{H}_0(\bs{k})+\mathcal{H}^{\textrm{SOC}}_{\textrm{int}}+\mathcal{H}^{\textrm{Z}} 
\label{eq:5Zeeman}
\end{equation}
with
\begin{equation}
\mathcal{H}^{\textrm{Z}} = B_1 \, \sigma_1 \rho_0 \tau_0 + B_2 \, \sigma_2 \rho_0 \tau_0 + B_3 \, \sigma_3 \rho_0 \tau_0
\label{eq:6Zeeman explicit}
\end{equation}
representing the Zeeman term.
Similarly to the case of $\mathcal{H}^{\textrm{SOC}}_{\textrm{R}}$, $\mathcal{H}^{\textrm{Z}}$ also splits into two components: the one parallel to the intrinsic SOC, and the other perpendicular to it.
In Fig.~\ref{fig:supple2qwe}, we present the numerical result of Eq.~\eqref{eq:6} using Hamiltonian Eq.~\eqref{eq:5Zeeman} in both cases for two different edge terminations, i.e., zigzag and armchair.
We observe that the perpendicular Zeeman field lifts the Dirac crossing of the topological boundary state, whereas the parallel one twists the edge states into an asymmetric form in $\boldsymbol{k}$-space, preserving the Dirac crossing.
The gaplessness of the edge states in the latter case is protected by the (artificial) spin-conservation along the axis of the magnetic field.

\subsection{Comparison of GGA and HSE06 for the band structure calculation}
\begin{figure}[h]
\centering
\includegraphics[width=0.26\textwidth]{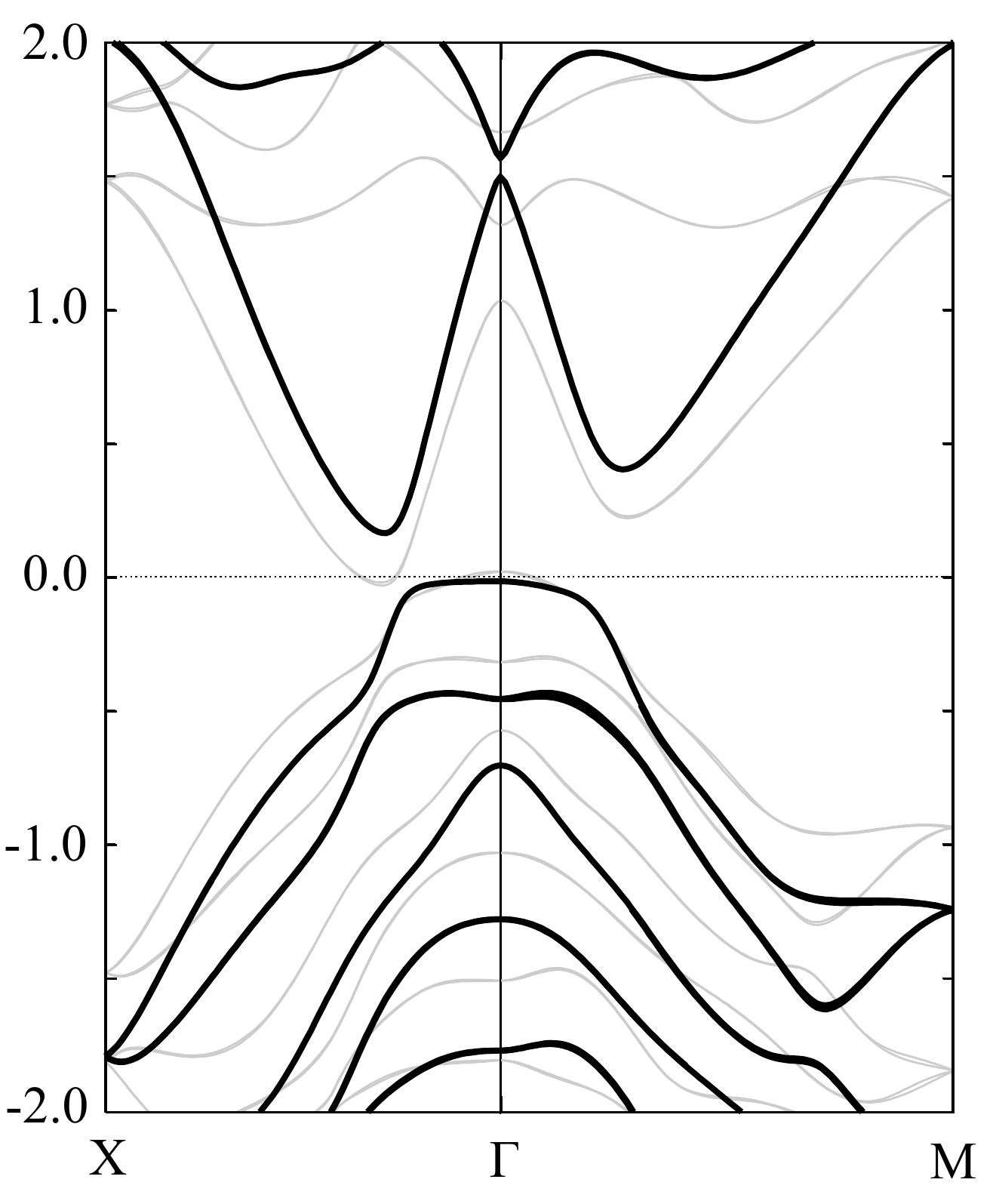}
\caption{Band structure of monolayer WTe$_2$ with SOC calculated using GGA (gray lines) and HSE06 (thick black lines) as exchange-correlation functionals.}
\label{fig:supple1qwe}
\end{figure}

In Fig.~\ref{fig:supple1qwe} we present the direct comparison between the band structures obtained from GGA and HSE06 with a Hartree-Fock exchange of $\alpha_{\mathrm{exx}}=0.25$. It is well known that GGA may over-emphasizes metallic screening effects due to an underestimate of band gaps. The inclusion of direct exchange mitigates this effect by reducing the self-interaction errors and leads to a band gap more in line with the experimental results.

\end{widetext}

\end{document}